\documentclass[aps,prl,superscriptaddress,reprint]{revtex4-1} 

\usepackage[caption=false]{subfig}
\usepackage{graphicx}
\usepackage{dcolumn}
\usepackage{bm}
\usepackage{epstopdf}
\usepackage{hyperref}

\begin{document}

\title{Pure Single Photons from Scalable Frequency Multiplexing}

\author{T. Hiemstra}
\thanks{Both authors contributed equally to this work.}
\affiliation{Clarendon Laboratory, Department of Physics, University of Oxford, Oxford OX1 3PU, United Kingdom}

\author{T.F. Parker}
\thanks{Both authors contributed equally to this work.}
\affiliation{Blackett Laboratory, Imperial College London, London SW7 2AZ, United Kingdom}
\affiliation{Clarendon Laboratory, Department of Physics, University of Oxford, Oxford OX1 3PU, United Kingdom}

\author{P. Humphreys}
\affiliation{Clarendon Laboratory, Department of Physics, University of Oxford, Oxford OX1 3PU, United Kingdom}

\author{J. Tiedau}
\affiliation{Clarendon Laboratory, Department of Physics, University of Oxford, Oxford OX1 3PU, United Kingdom}

\author{ M. Beck}
\affiliation{Department of Physics, Reed College, Portland, Oregon 97202, USA}

\author{ M. Karpi\'{n}ski}
\affiliation{Clarendon Laboratory, Department of Physics, University of Oxford, Oxford OX1 3PU, United Kingdom}
\affiliation{Faculty of Physics, University of Warsaw, Pasteura 5, 02-093 Warszawa, Poland}
 
\author{B.J. Smith}
\affiliation{Clarendon Laboratory, Department of Physics, University of Oxford, Oxford OX1 3PU, United Kingdom}
\affiliation{Department of Physics and Oregon Center for Optical, Molecular, and Quantum Science, University of Oregon, Eugene, Oregon 97403, USA}

\author{ A. Eckstein}
\affiliation{Clarendon Laboratory, Department of Physics, University of Oxford, Oxford OX1 3PU, United Kingdom}
 
\author{W.S. Kolthammer}
\email[]{Corresponding author: steve.kolthammer@imperial.ac.uk}
\affiliation{Blackett Laboratory, Imperial College London, London SW7 2AZ, United Kingdom}
 
\author{ I.A. Walmsley}
\affiliation{Blackett Laboratory, Imperial College London, London SW7 2AZ, United Kingdom}
\affiliation{Clarendon Laboratory, Department of Physics, University of Oxford, Oxford OX1 3PU, United Kingdom}













\begin{abstract}
We demonstrate multiphoton interference using a resource-efficient frequency multiplexing scheme, suitable for quantum information applications that demand multiple indistinguishable and pure single photons. In our source, frequency-correlated photon pairs are generated over a wide range of frequencies by pulsed parametric down conversion. Indistinguishable single photons of a predetermined frequency are prepared using frequency-resolved detection of one photon to control an electro-optic frequency shift applied to its partner. Measured photon statistics show multiplexing increases the probability of delivering a single photon, without a corresponding increase to multiphoton events. Interference of consecutive outputs is used to bound the single-photon purity and demonstrate the non-classical nature of the emitted light. 
\end{abstract}


\maketitle
\section{Introduction}
The deterministic preparation of an optical field consisting of exactly one photon with completely specified characteristics is a long-running scientific and technical challenge \cite{eisaman_invited_2011}. Potential applications for such a light source include sensing and imaging, quantum computing and simulations, secure communications, and fundamental tests of quantum science \cite{ZeilingerReview2005, OBrienReview2009, WalmsleyReview2015}. Approaches to meet this challenge are predominantly based on either a single emitter or nonlinear optical wave mixing, and the performance of such a single-photon source may be characterized in terms of its photon statistics-- the degree to which the output field contains precisely one photon, and the single-photon purity-- the degree to which the single photon occupies a single optical space-time mode. For some applications specific modal characteristics (e.g. particular frequency and beam geometry) might also be required. In all of these aspects, significant progress has been made for both single-emitter and nonlinear optical sources \cite{Senellart2017,Aharonovich2016, Flamini_PhotonicQIReview_2018}, yet further improvements are necessary to enable the most demanding applications.

Nonlinear optical photon sources are based on heralding detection and spontaneous parametric wave mixing, most commonly spontaneous parametric down conversion (SPDC). In SPDC the second-order nonlinear response of a material due to its interaction with a strong pump field generates pairs of photons in two new fields, which we refer to as signal and herald. Detection of the herald field is used to indicate when a single photon is prepared in the signal field \cite{Zeldovich1969,ChristReview2013}.

A quantum mechanical description of SPDC is given by an effective Hamiltonian that is bilinear in field operators:
\begin{equation}
\hat{H}=B \int d\nu\,d\nu'\,f\left(\nu,\nu'\right)\hat{a}^\dagger_\mathrm{s}\left(\nu\right) \hat{b}^\dagger_\mathrm{h}\left(\nu'\right)+\mathrm{h.c.}
\label{eq:H}
\end{equation}
where $\hat{a}^\dagger_\mathrm{s}$ and $\hat{b}^\dagger_\mathrm{h}$ are bosonic creation operators in the signal and herald fields at the specified frequency $\nu$, and the gain $B$ depends on the optical nonlinearity and pump field intensity. The complex-valued joint spectral amplitude (JSA) function $f\left(\nu,\nu'\right)$ describes spectral-temporal correlations between the signal and herald fields \cite{Grice_PDC}.

\begin{figure}[t]
    \includegraphics[width=1.\linewidth]{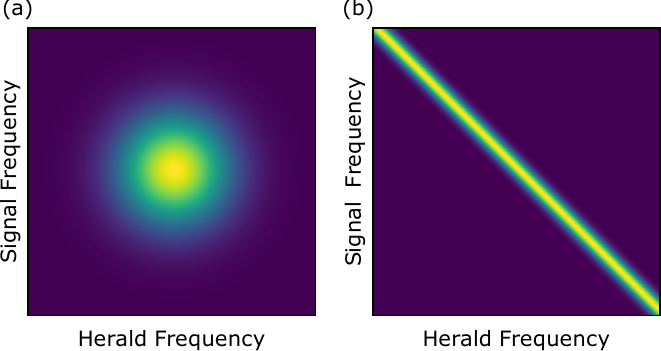}
	\caption{A joint spectral intensity $|f(\nu, \nu')|^2$ that is (a) factorable and (b) highly anticorrelated.}
	\label{fig:jointspectra}
\end{figure}

To achieve a high single-photon purity, heralded sources often employ a JSA that is factorable, $f\left(\nu,\nu'\right)=A_\mathrm{s}(\nu)B_\mathrm{h}(\nu')$, as in Fig.~\ref{fig:jointspectra}(a) \cite{Grice2001}. The frequency characteristics of the signal photon are therefore independent of those of the detected herald photon, and Eq.~\ref{eq:H} results in a two-mode squeezed vacuum state in broadband modes $A_\mathrm{s}$ and $B_\mathrm{h}$ with a squeezing parameter dependent on the gain. In this case, with optimal squeezing, the maximum probability to generate one pair of photons is 0.25. If the heralding detectors are capable of efficient photon number resolution, this allows such a high gain source to be employed. However, typically such detectors are at best inefficient, and more usually are not photon number resolving, and therefore a common expedient is to suppress contributions from multiple photon pairs, by operating the single-photon sources far below this limit \cite{Christ_limits_2012}. 

To increase the probability of delivering a photon, the outputs from several heralded SPDC sources can be combined into a common channel, an approach referred to as source multiplexing \cite{pittman_single_2002,migdall_tailoring_2002}. A spatially multiplexed photon source, for example, consists of multiple factorable sources routed through a switching network, conditioned on which source generates a photon pair \cite{migdall_tailoring_2002}. In a frequency-multiplexed photon source \cite{grimau_puigibert_heralded_2017, joshi_frequency_2018}, the spectral structure of a single nonlinear interaction plays the role of multiple factorable sources. By using an anticorrelated JSA (Fig. \ref{fig:jointspectra}(b)), which naturally arises when the pump bandwidth is much less than the phase-matching bandwidth, photon pairs that differ only by their central frequencies are generated across the whole joint spectrum. The frequency of a signal photon is determined by a frequency-resolved heralding measurement. A frequency shift is then applied to route the signal photon to the specified output mode, matched to the passband of an output filter needed to reject photons at other frequency ranges.

Recent advances in both single-emitter and multiplexed single-photon sources have led to devices that achieve a 50-70\% probability of delivering a single photon with a purity of 80-90\%  \cite{Kaneda_time-2018, Senellart2017}. Multiplexed devices could be further improved by employing number resolved herald detection-- an ideal device would then require only 17 heralded sources to deliver a photon with over 99\% probability \cite{Christ_limits_2012}, however to be practical this would require a substantial improvement to the achievable detection efficiency. An alternative approach that makes use of efficient, but non-number-resolving detectors, is to use a large number of sources of weakly-squeezed light, where the probability of more than one pair of photons being generated is very small.

The technical feasibility for a multiplexing scheme to employ a large number of sources depends on how the number of device components and overall loss scale with the number of sources $N$. Spatial schemes require $N$ physically distinct SPDC sources, and the construction of the $N\times 1$ switch determines its loss scaling: For example, a binary tree network requires log N two-port switches; a generalised Mach-Zehnder network demands N couplers and a phase shifter (used X times) and a serial network N switches. \cite{Bonneau_multLoss_2015, Gimeno_Segovia_RMUX_2017}. In contrast, temporal multiplexing with an optical delay requires only one two-port switch and one SPDC source,each operated $N$ times \cite{Kaneda_time-2018}. As with serial spatial switches, however, the output suffers a worst-case loss of $N$ passes through the switch. Frequency multiplexing is unique in combining advantages of the $N \times 1$ switch and optical delay. In this case, neither the overall loss nor the number of physical components necessarily increases with $N$.

In this article, we report a frequency multiplexing device that demonstrates the promise to scale to high performance using a large number of effective sources. In particular, the methods we use to generate, detect, and switch photons operate continuously over a range of frequencies, thereby realizing the scaling advantage of this approach. We demonstrate an enhancement in single-photon statistics and, for the first time, test the single-photon purity of a frequency multiplexed source through interference of two photons generated by the source on sequential trials.. 

\section{Experimental Methods}
\begin{figure}[b!]
    \includegraphics[width=1.\linewidth]{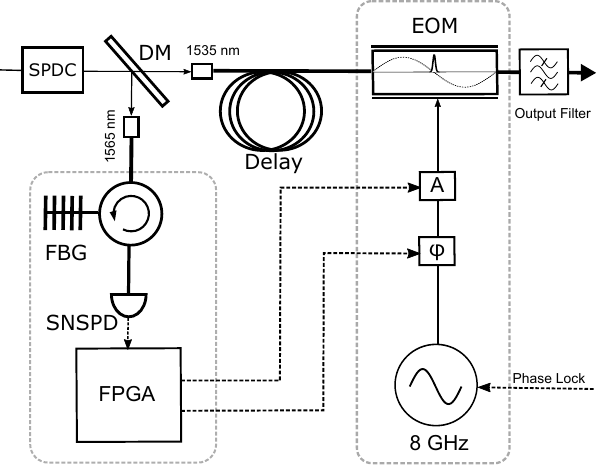}
	\caption{Schematic of frequency multiplexed source. Spontaneous parametric down-conversion (SPDC) generates frequency correlated fields which are separated by a dichroic mirror (DM). Measurement of the long-wavelength herald field by a time-of-flight spectrometer, comprised of a chirped fiber Bragg grating (FBG) and superconducting nanowire single-photon detector (SNSPD), is used to determine the frequency shift applied to the short-wavelength signal field by an electro-optic phase modulator (EOM).}
	\label{fig:setup}
\end{figure}

A schematic of our frequency-multiplexed single-photon source is shown in Fig.~\ref{fig:setup}. Pairs of photons with anticorrelated frequencies are generated by pulsed type-0 SPDC in a periodically poled potassium titanyl phosphate waveguide. This process is driven by a pulsed pump field at 386.8~THz (775~nm) with a full width half maximum (FWHM) bandwidth of 60~GHz (0.12~nm) and repetition rate of 10~MHz. The resulting joint spectrum is similar to that shown in Fig.\,\ref{fig:jointspectra}(b), symmetric about 193.5~THz (1550~nm) and highly linear over a range of more than 10~THz (80~nm). The herald and signal fields, respectively corresponding to wavelengths longer and shorter than 1550~nm, are divided in two beams by a dichroic mirror.

The herald field is then measured by a time-of-flight single-photon spectrometer \cite{avenhaus_fiber-assisted_2009, davis_pulsed_2016} consisting of a chirped fibre Bragg grating (FBG) and time-resolved detection by a superconducting nanowire single-photon detector (SNSPD). The FBG results in a frequency-dependent delay of 16~ps/GHz. The detection time, recorded by an FPGA-based time-to-digitial converter (TDC) \cite{Bourdeauducq_TDC}, determines the frequency of a herald photon with an uncertainty of 10~GHz due jitter in the SNSPD and timing electronics. 

The frequency of a heralded photon in the signal field is shifted using a travelling-wave electro-optic phase modulator (EOM) \cite{cumming_serrodyne_1957, Wright_2017_spectralShear}. The photon co-propagates through the EOM with a drive voltage $V(t)=V_0 \sin \left( 2\pi\nu_\mathrm{RF}t \right)$ with frequency $\nu_\mathrm{RF} = 8\mathrm{~GHz}$ and amplitude $V_0$. The period is substantially longer than the photon's pulse duration, so that appropriately setting the drive phase with respect to the photon's arrival time causes a phase across the single-photon pulse that varies linearly in time. The corresponding shift in the photon's carrier frequency is  
\begin{equation}\label{eqn:shift}
 \Delta\nu =\pi\frac{V_0}{V_{\pi}} \nu_\mathrm{RF}.
\end{equation}
where $V_\pi$ is the EOM voltage that results in an optical phase of $\pi$.
\begin{figure}[t]
	\includegraphics[width=1.\linewidth]{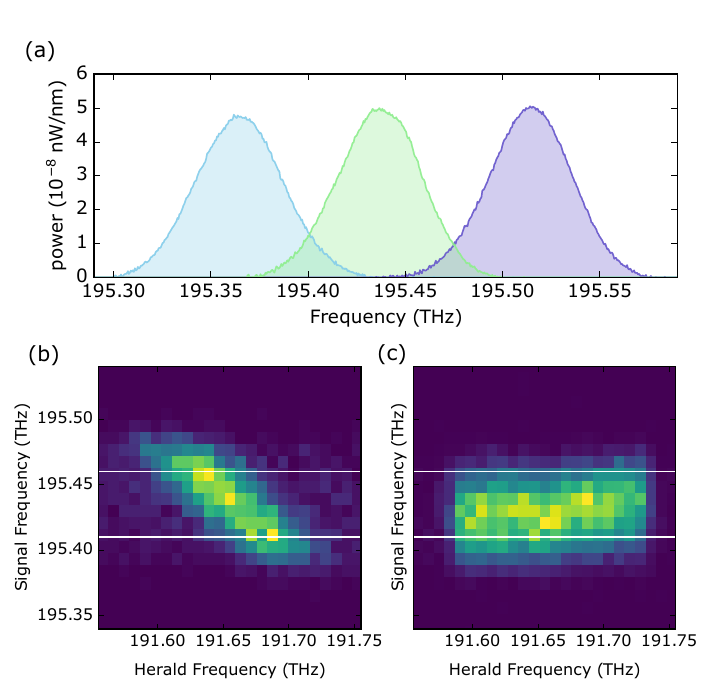}%
	\caption{Demonstration of frequency shifting. (a) Spectrum of the signal field, generated by difference frequency generation, after maximal negative (blue), no (green) or maximal positive (purple) frequency shift. (b-c) The joint spectrum of a photon pair (b) without shifting and (c) with shifting conditioned on the measured herald frequency. The white lines indicate the 50 GHz passband of the signal output filter.}
	\label{fig:shifting}
\end{figure}
To calibrate the frequency shifting, a continuous-wave seed field is added at 1565~nm, so that signal field pulses generated via difference frequency generation can be readily measured with an optical spectrum analyzer. Figure~\ref{fig:shifting}(a) demonstrates a total frequency shift range of $2 \Delta\nu_\mathrm{max}$ = 170 GHz using this method. The shifted spectral distributions show greater than 94\% overlap, indicating that the shift causes negligible distortion.     

Lastly, we combine the frequency-resolved herald measurement and frequency shift using feed-forward control. The FPGA uses a calibrated look up table to set the EOM voltage $V_0$ with a voltage-controlled attenuator, based on the time-of-flight detection time. To measure the joint spectrum of photon pairs, a second time-of-flight spectrometer is connected to the output signal field. Figure \ref{fig:shifting}(b) shows that without shifting, the joint spectrum exhibits anticorrelation as expected. The signal output filter, with a passband indicated by the white lines, limits the observed frequency range. In contrast, the data in Fig.~\ref{fig:shifting}(c) correspond to active frequency shifting. The joint distribution now shows that the frequency of the signal photon is largely independent of the measured frequency of the herald photon.

\section{Source Characterisation}
\begin{figure}[b!]
	\includegraphics[width=1\linewidth,height=1.2\linewidth]{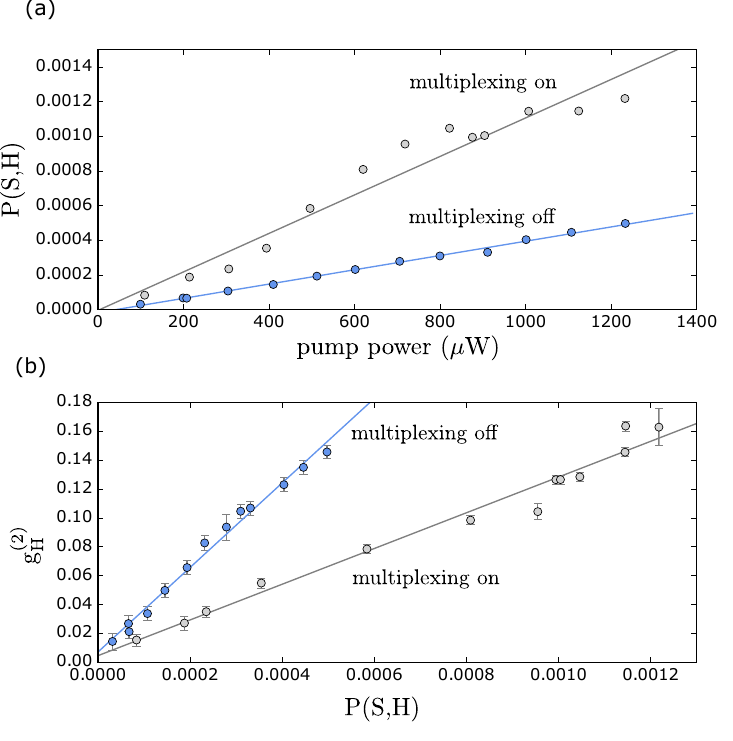}%
	\caption{\label{fig:gH2_cc} Measurement of photon statistics with multiplexing inactive (blue) and active (grey). As the pump power is varied, we measure (a) the probability of a joint detection event in the signal and herald fields $P(\mathrm{S},\mathrm{H})$ and (b) the second-order correlation $\mathrm{g}_{\mathrm{H}}^{(2)}$ in the case of a herald detection event.}
	\label{fig:statistics}
\end{figure}

We first examine how much the probability of delivering a photon increases when frequency shifting is activated. To do so, we measure the probability $P\left(\mathrm{S},\mathrm{H}\right)$ of joint detection events by SNSPDs on the output signal and herald fields. Figure~\ref{fig:gH2_cc}(a) shows the dependence of $P\left(\mathrm{S},\mathrm{H}\right)$ on the average power, with and without shifting activated. In both cases, we observe that $P\left(\mathrm{S},\mathrm{H}\right)$ increases linearly with power, as expected in the limit of small squeezing parameters. For a given pump power, we measure $P\left(\mathrm{S},\mathrm{H}\right)$ to increase by a factor of 2.7 when multiplexing is engaged.

Furthermore, we show that this increase in joint detection probability is achieved without a degredation of the output's single-photon character, as is the case if the pump power is simply increased without using multiplexing. For this we measure the heralded second-order correlation $\mathrm{g}_{\mathrm{H}}^{(2)}$ using a Hanbury-Brown-Twiss (HBT) setup of a beam splitter followed by two detectors on the output of the source. As above, we record herald H and signal S$_1$ and S$_2$ detection events and estimate their independent and joint probabilities. We then estimate the second-order correlation from 
\begin{equation}
\mathrm{g}_{\mathrm{H}}^{(2)}  = 
\frac{\langle n^2\rangle-\langle n\rangle}{\langle n\rangle^2}
\approx\frac{P\left(\mathrm{S_1},\mathrm{S_2},\mathrm{H}\right)P\left(\mathrm{H}\right)}{P\left(\mathrm{S_1},\mathrm{H}\right)P\left(\mathrm{S_2},\mathrm{H}\right)}
\end{equation}
where $\langle n\rangle$ is the mean number of photons in the signal field, before the HBT setup, conditional on a herald detection event, and in the second step we make an approximation of small squeezing \cite{U'ren_Characterization_2005}. For an ideal single-photon source $\mathrm{g}_{\mathrm{H}}^{(2)}=0$ since there is no probability to deliver more than one photon. For an ideal non-multiplexed heralded source with small squeezing, $\mathrm{g}_{\mathrm{H}}^{(2)}=4p$ for probability $p$ to generate a pair of photons \cite{Belhassen}. 

Figure~\ref{fig:gH2_cc}(b) shows $\mathrm{g}_{\mathrm{H}}^{(2)}$ measured for different values of $P\left(\mathrm{S},\mathrm{H}\right)$, which are achieved by varying the pump power as indicated in Fig.~\ref{fig:gH2_cc}(a). Without multiplexing, an increase in $P\left(\mathrm{S},\mathrm{H}\right)$ is accompanied by an undesirable increase in $\mathrm{g}_{\mathrm{H}}^{(2)}$. By using multiplexing, these data show an increase in the probability of delivering a photon is achieved while maintaining a constant $\mathrm{g}_{\mathrm{H}}^{(2)}$.  

\begin{figure}[!htbp]
	\includegraphics[width=.9\linewidth]{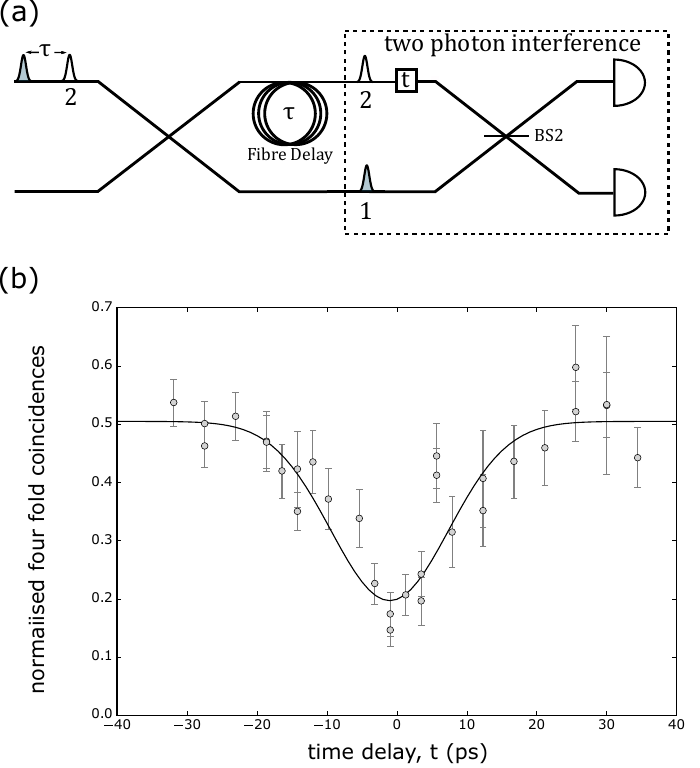}
	\caption{Interference of two single-photon outputs. (a) Experimental schematic consisting of an initial beam splitter (BS1), fixed delay $\tau$ and adjustable delay $t$, interference beam splitter (BS2), and two single-photon detectors. (b) The rate of coinciding detection events for varying delay $t$,  with the black line to guide the eye. The visibility of the interference dip is 0.61 $\pm$ 0.04}
	\label{sfig:HOMsetup}
\end{figure}

We now investigate the modal purity of photons delivered by the multiplexed source \cite{Christ_multimode_corr_2011}. In particular, the spectral mode should be consistent and independent of the herald frequency measurement. The modal purity is estimated by interfering consecutive heralded photons \cite{Moslet_purity_2008} using an unbalanced Mach-Zehnder interferometer. As shown in Fig. \ref{sfig:HOMsetup}(a) the two photons are probabilistically separated by an initial beam splitter and aligned in time by a fixed optical-fiber delay and adjustable micro-positioned delay. Interference at a second beam splitter is observed with single-photon detectors monitoring each output.

Two-photon interference is observed through variations in the joint detection probabilities as the relative delay is changed \cite{hong_measurement_1987}, as shown in Fig.~\ref{sfig:HOMsetup}(b). These data are normalised to account for fluctuations in the pump power, causing a varying photon-pair generation rate, between each data point. This is achieved by dividing the raw fourfold coincidence counts by the square of the total heralded signal counts.

The maximum state purity is inferred to be the visibility of this variation in joint detection \cite{Moslet_purity_2008}. We observe an interference visibility of 0.61 $\pm$ 0.04, noting that a visibility of over $0.5$ from phase-independent sources indicates non-classical light \cite{Mandel1983}.

The modal purity of the single photons can be inferred to be at least as large as the observed interference visibility \cite{Moslet_purity_2008}. A detailed model of our source indicates the purity
suffers two technical limitations specific to its implementation. First, limited resolution of the herald spectrometer, due to timing jitter in detection electronics, limits the purity to 0.92. Second, group velocity dispersion in the optical delay line (300~m of SMF-28 fiber) that precedes frequency shifting causes a time delay correlated to measured herald frequency. The combined effects of spectrometer uncertainty and dispersion result in a purity of 0.84. We estimate that further reductions in the two-photon interference visibility are due to multi-pair generation and imperfections in our interferometer. These calculations are presented in detail in the Supplementary Material.

\section{Outlook}

In order to maximise the probability of delivering a single photon it is necessary to address as many frequency modes as possible whilst also minimising the losses on the signal and herald fields (see Supplementary Material for a full discussion about losses). For a continuous joint spectrum, the number of accessible modes is proportional to the ratio of the shift range to the photon bandwidth, $2 \Delta \nu_{\mathrm{max}} / \sigma_{\mathrm{photon}} $. For a sinusoidal signal driving an EOM, the period of the drive signal must be larger than the photon temporal bandwidth in order to ensure a linear phase is applied to the photon. For an optimal ratio between the drive signal period and the photon temporal bandwidth, the maximum mode number is proportional to $\mathrm{V_0}$ / $\mathrm{V}_{\pi}$. 

The SPDC source described already generates photon pairs with anticorrelated frequencies that span a range over 80 nm. The range of the single-photon spectrometer described is 27 times greater than used. The limitation in our current implementation comes from the maximum rf voltage applied to the EOM. We expect that this can be increased by more than 30 times using standard RF amplifiers up to the breakdown voltage of our current EOM.

The single-photon purity can be readily improved by minimizing dispersion in the optical delay line. Splicing a suitable length of dispersion-compensating  into the delay line, for example, would achieve this with negligible additional attenuation. Impurity due to heralding uncertainty can be reduced by increasing the spectrometer resolution, either by using low-jitter SNSPDs or increasing the FBG optical dispersion in the time-of-flight detector. Alternately, the spectrometer requirements could be greatly reduced by using a cavity-based SPDC source designed to generate photon pairs in distinct factorable states comprised of two longitudinal cavity modes \cite{Jeronimo-Moreno2010, Luo_2015}.

The single-photon spectrometer and frequency shifter described here are integrated using standard optical fiber. A complete waveguide-integrated multiplexed source can be achieved by employing a photon-pair source based on a fiber-coupled photonic chip \cite{PhysRevApplied.8.024021} or in-fiber nonlinear optics \cite{Francis-Jones:16}, along with a fiber wavelength splitter. The resulting device would allow compact and robust alignment-free operation. In light of the feasibility and performance scaling investigated here, these practical considerations suggest frequency multiplexing is a promising route to meeting the long-running challenge of a near-deterministic source of highly pure single photons. 

\section*{Funding Information}

Engineering and
Physical Sciences Research Council (NQIT EP/M013243/1 and
BLOQS EP/K034480/1); European Research Council (MOQUACINO); TFP is supported by the EPSRC Centre for Doctoral Training in Controlled Quantum Dynamics (EP/L016524/1).

\section*{Acknowledgments}
The authors thank Alex Davis, Helen Chrzanowski and Ben Metcalf for fruitful discussions.

\section*{Supplemental Documents}
See Supplement 1 for supporting content.


\bibliography{bib}


\clearpage

\end{document}


\title{ Supplementary Information }



\begin{center}
\textbf{Supplementary Information}
\end{center}
\section{Experimental Details}

A frequency doubled telecom fibre laser and EDFA system is used to generate a pump field with a repetition rate of 10 MHz, a bandwidth of 0.12 nm and a central frequency of 775 nm.
 
The photon pair source is based on spontaneous parametric down-conversion (SPDC) in periodically poled potasium titanyl phosphate (KTP). We use a KTP waveguide of length 3.2 mm, supplied by ADVR. A type-0 phase matching process results in co-propagating photon pairs for which both photons have the same polarization as the and pump field. After the KTP waveguide, the pump is removed via filtering and photon pairs are separated on a dichroic mirror. The dichroic mirror reflects wavelengths larger than 1550\,nm, while transmitting short wavelengths. This cutoff wavelength is matched to the degeneracy point of the SPDC. Each photon pair is therefore separated into two spatial modes, one containing heralding photons centred at 1565\,nm and the other containing signal photons centred at 1535\,nm.

The time of flight (TOF) spectrometer uses a time to digital converter (TDC) based on a Spartan-6 FPGA with modified software developed by \cite{Bourdeauducq_TDC}. This system allows us to measure the herald photon time relative to a clock signal derived from the 10 MHz laser system with a resolution of up to 33\,ps. This, combined with the fibre Bragg grating (with a dispersion of -2.3\,ns/nm) allows us to measure the  frequency of the herald photon with a resolution of approximately 10\,GHz. A look up table (LUT) contains a mapping from the arrival time of the herald photon (and therefore its frequency) to the voltage settings that need to be applied to a RF attenuator and phase shifter in order to apply the correct RF drive signal to the phase modulator.

\section{Purity of Frequency Multiplexed Photons}

It is important that single photons from a frequency multiplexed source are in highly pure states \cite{Senellart2017}. Imperfect frequency detection, resulting from timing jitter on the photon detectors and timing electronics, group velocity dispersion in delay stage before frequency shifting, and jitter on the EOM drive signal all contribute to decrease the purity of single photons delivered by the source. Here we describe how these influence the single photon purity, and then calculating the expected purity based on the parameters of our experiment.

The state of a heralded signal photon, following a heralding measurement outcome of frequency $\omega_H$, is given by

\begin{equation}
\begin{split}
\boldsymbol{\rho}_{\omega_\mathrm{H}} = & \int d\omega_\mathrm{i} P(\omega_\mathrm{i}|\omega_\mathrm{H})\\
& \cdot \int \int d\omega_\mathrm{s} d\omega_\mathrm{s'} f(\omega_\mathrm{s}, \omega_\mathrm{i}) f^*(\omega_\mathrm{s'}, \omega_\mathrm{i}) |\omega_\mathrm{s}\rangle\langle\omega_\mathrm{s'}|
\end{split}
\end{equation}
where $f(\omega_\mathrm{s}, \omega_\mathrm{i})$ is the joint spectral amplitude of a photon pair. In general, the joint spectral amplitude describing photons generated by the SPDC consists of a product of a pump envelope function and a phase matching function for the non linear material \cite{Grice_PDC}. For our experiment, in the frequency region in which we herald photons, the phase matching function is much broader than the pump envelope function. The joint spectral amplitude can therefore be well approximated by the pump envelope function alone. 

The conditional probability distribution $ P(\omega_\mathrm{i}|\omega_\mathrm{H}) $ describes the probability of the herald photon having a frequency $\omega_\mathrm{i}$ given that a measurement outcome of frequency $\omega_\mathrm{H}$. This distribution characterises the imperfect frequency detection of the time of flight spectrometer. 

After heralding, the signal photons travel through a delay stage in order to compensate for the latency of the feed forward electronics. Each heralded photon obtains a frequency dependant phase in the delay line before frequency shifting as a result of chromatic dispersion. A plane wave propagating through a dispersive medium obtains the phase $k(\omega) L$, where $k(\omega)$ can be expanded about $\omega_\mathrm{0}$ so that 

\begin{equation}
\begin{split}
k(\omega) = &
k(\omega_\mathrm{0}) + k'(\omega_\mathrm{0})(\omega - \omega_\mathrm{0}) +\\ &\frac{1}{2}k''(\omega_\mathrm{0})(\omega - \omega_\mathrm{0})^2 + O(\omega^3).
\end{split}
\end{equation}

We take into account terms up to the quadratic term, which describes a phase acquired by each signal photon $\gamma(\omega_\mathrm{s}  - \omega_\mathrm{0})^2$, where we introduce a group velocity dispersion (GVD) parameter $\gamma = \frac{1}{2}k''(\omega_\mathrm{0}) L$. This phase corresponds to a time delay to each photon after the delay line that depends on its frequency. 

After the delay line, the signal photon is frequency shifted to a central target frequency $\omega_\mathrm{c}$ (the central frequency of the output filter) by the application of a linear temporal phase, the exact form of which is set by the frequency of the herald measurement. In our experiment this is achieved using a sinusoidal phase, where the signal photon wave packet is phase locked to the linear region, given by

\begin{equation}
\phi(t) = \frac{(\omega_c - \omega_H + \omega_p)}{\omega_{d}} \mathrm{sin}(\omega_{d} t) \approx (\omega_c - \omega_H -\omega_p) t 
\end{equation}

Finally, signal photons pass through an output filter with central frequency $\omega_c$.  This is well approximated by a top-hat filter $F(\omega_s)$  on the signal mode with a bandwidth of $\sigma_f$.

By taking these processes into account, we can write down the form of the overall state generated by the frequency multiplexing experiment as

\begin{equation}
\begin{split}
\boldsymbol{\rho} = &\int d \omega_\mathrm{H} P(\omega_\mathrm{H}) \int d \omega_\mathrm{i} P(\omega_\mathrm{i} | \omega_\mathrm{H})\\
&\cdot \int\int d\omega_\mathrm{s} d \omega'_\mathrm{s} A(\omega_\mathrm{s}, \omega_\mathrm{i}| \omega_\mathrm{H}) A^*(\omega'_\mathrm{s}, \omega_\mathrm{i} | \omega_\mathrm{H})|\omega_\mathrm{s}\rangle \langle\omega'_\mathrm{s}|
\end{split}
\end{equation}

The function $A(\omega_\mathrm{s}, \omega_\mathrm{i}| \omega_\mathrm{H})$ is the signal photon wave packet, conditioned on measuring the herald photon at frequency $\omega_\mathrm{H}$ and is given by

\begin{equation}
\begin{split}
A(\omega_\mathrm{s}, \omega_\mathrm{i} | \omega_\mathrm{H}) =& N(\omega_\mathrm{H}, \omega_\mathrm{i}) F(\omega_\mathrm{s})\exp\left[\frac{-(\omega_\mathrm{s} + \omega_\mathrm{h} - \omega_\mathrm{i} +\omega_\mathrm{c} )^2}{2\sigma^2}\right]\\ & \times \exp\left[i\gamma(\omega_\mathrm{0} + \omega_\mathrm{c}- \omega_\mathrm{d} + \omega_\mathrm{H} + \omega_\mathrm{s})^2 \right]
\end{split}
\end{equation}
where $N(\omega_\mathrm{H}, \omega_\mathrm{i})$ is a normalisation constant, so that $\int d\omega_\mathrm{s}|A(\omega_\mathrm{s}, \omega_\mathrm{i} | \omega_\mathrm{H})|^2 = 1$. 
The purity of this state is given by

\begin{equation}
\begin{split}
Tr(\boldsymbol{\rho}^2) =& \int\int d \omega_\mathrm{H} d\omega'_\mathrm{H} P(\omega_\mathrm{H}) P(\omega'_\mathrm{H}) \\
&\cdot \int \int d \omega_\mathrm{i} d \omega'_\mathrm{i} P(\omega_\mathrm{i} | \omega_\mathrm{H}) P(\omega'_\mathrm{i} | \omega'_\mathrm{H})  \\
&\cdot \left| \int  d\omega_\mathrm{s} A(\omega_\mathrm{s}, \omega_\mathrm{i}| \omega_\mathrm{H}) A^*(\omega_\mathrm{s}, \omega'_\mathrm{i}| \omega'_\mathrm{H})\right|^2.
\label{PurityEq}
\end{split}
\end{equation}

\subsection{Calculating the Purity of Frequency Multiplexed Photons}

Both the imperfect nature of the frequency resolved detection and the group velocity dispersion experienced by heralded single photons in the delay line act to reduce the state purity of the frequency multiplexed source. We now calculate the expected purity of the photons from our experiment using equation \ref{PurityEq} and the experimental parameters summarised in table  \ref{experimentParams1}.

\begin{table}[h]

\begin{tabular}{c|c}

\hline
Parameter   & Value   \\

\hline
$\sigma_{\textrm{pump}}$      & $2 \pi \cdot 50.95  \times 10^9$ rad $\cdot$ s$^{-1}$       \\
$\sigma_{\textrm{filter}}$       & $2\pi$  $\cdot50$ $\times$  $10^9$ rad $\cdot$ s$^{-1}$          \\
$\omega_\textrm{d}$    & $2\pi c / 775  \times 10^9$ rad $\cdot$ s$^{-1}$ \\
$\omega_\textrm{c}$  &  $2\pi c / 1535 \times 10^9$ rad $\cdot$ s$^{-1}$  \\
$\gamma$    & $\leq$  3.5 $\times$ 10$^{-24}$ s$^{2}$           \\
$\Delta \omega_{\textrm{H}}$  & 1.11 $\times$ 10$^{12}$ rad $\cdot$ s$^{-1}$ \\

\end{tabular}

\caption{Experimental parameters and corresponding values. }
\label{experimentParams1}
\end{table}

\subsubsection{Characterisation of Imperfect Detectors}

\begin{figure}[htb!]
	\centering
		\includegraphics[width=0.5 \textwidth]{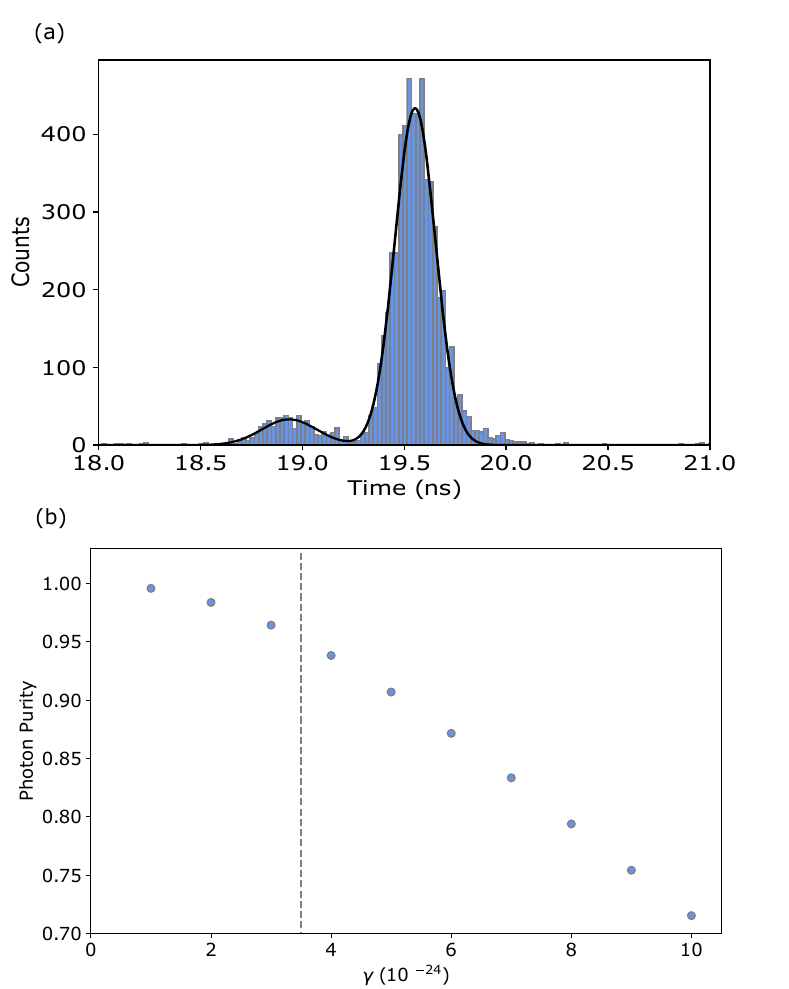}
		\caption{a) Histogram of timing jitter in the TOF spectrometer. (b) The calculated single photon purity as a function of the GVD parameter $\gamma$, for perfect frequency resolution on the measurement of the herald photon.}
        \label{jitterProb1}
\end{figure}

The frequency resolved detector is comprised of a highly dispersive fibre Bragg grating (FBG), which maps the frequency of a photon onto its arrival time at a single photon detector. The arrival time is measured relative to a clock signal, derived from the pump laser, using an FPGA (TDC). Timing jitter in the photon detectors and the TDC results in uncertainty in the photon frequency. This uncertainty is characterised by measuring the arrival time of photons generated at a well defined frequency. Figure \ref{jitterProb1} shows the arrival time measurements of photons generated using seeded SPDC to generate photons with a carrier frequency of 1535\,nm. The histogram bin width is defined by the TDC resolution of 33\,ps.  

We use a smooth fit to this distribution, along with the specified FBG dispersion, to approximate the probability distribution $P(\omega_\mathrm{H}| \omega_\mathrm{i})$. From this we approximate $P(\omega_{\mathrm{i}}$| $\omega_\mathrm{H})$ (Figure \ref{jitterProb1}). The isolated effect of imperfect frequency detection on the heralded photon purity is calculated from equation \ref{PurityEq} by taking $\gamma =0$. The resulting heralded photon purity is 0.92.  

\subsubsection{Group Velocity Dispersion}

The delay line is made from 300\,m of optical fiber (SMF\,28), with a specified maximum dispersion parameter of 18\,ps/(nm\,$\cdot$\,km). To investigate the effect of group velocity dispersion alone, we now assume perfect detector resolution, $P(\omega_\mathrm{i}|\omega_\mathrm{H}) = \delta(\omega_\mathrm{i} -\omega_\mathrm{H})$. 

Group velocity dispersion results in a time delay that depends on the herald frequency, which reduces the purity. This effect also acts to broaden the two photon interference curve we measure. Figure \ref{jitterProb1} shows the effect of GVD on the photon purity. For the specified maximum dispersion of 18 ps/(nm\,$\cdot$\,km) we calculate a resulting purity is 0.95, assuming perfect frequency detection.

\subsubsection{Overall Photon Purity}

We calculate that with the maximum expected GVD of 18 ps/(nm $\cdot$ km) and the measured timing jitter distribution we expect a minimum single photon purity of 0.84.

\subsection{Phase Jitter}

So far we have assumed a completely linear temporal phase is applied to each photon. Phase jitter in the microwave signal driving the phase modulator can result in an incorrect, and possibly non-linear, phase being applied.

In order to investigate this effect, we calculate the reduction in purity due to temporal jitter on the drive signal on a pure photon wave packet. In the time domain, after the application of the drive signal this is given by

\begin{equation}
A(t, x) = e^{-\frac{t^2\sigma^2}{2} + i(\omega_\mathrm{i} - \omega_\mathrm{d})}e^{i \frac{(\omega_\mathrm{c} - \omega_\mathrm{i}+\omega_\mathrm{d})}{\omega_{rf}} sin(\omega_\mathrm{{rf}}(t + x))}
\end{equation}
where we take $x$ to be a random variable from a Gaussian distribution with variance $\sigma_\textrm{jitter}^2$.

We then calculate the purity as
\begin{equation}
Tr(\rho^2) = \int \int dx dx' P(x) P(x') |\int A(t, x)A^*(t, x') dt|^2.
\end{equation}

\begin{figure}[htb!]
	\centering
		\includegraphics[width=0.5 \textwidth]{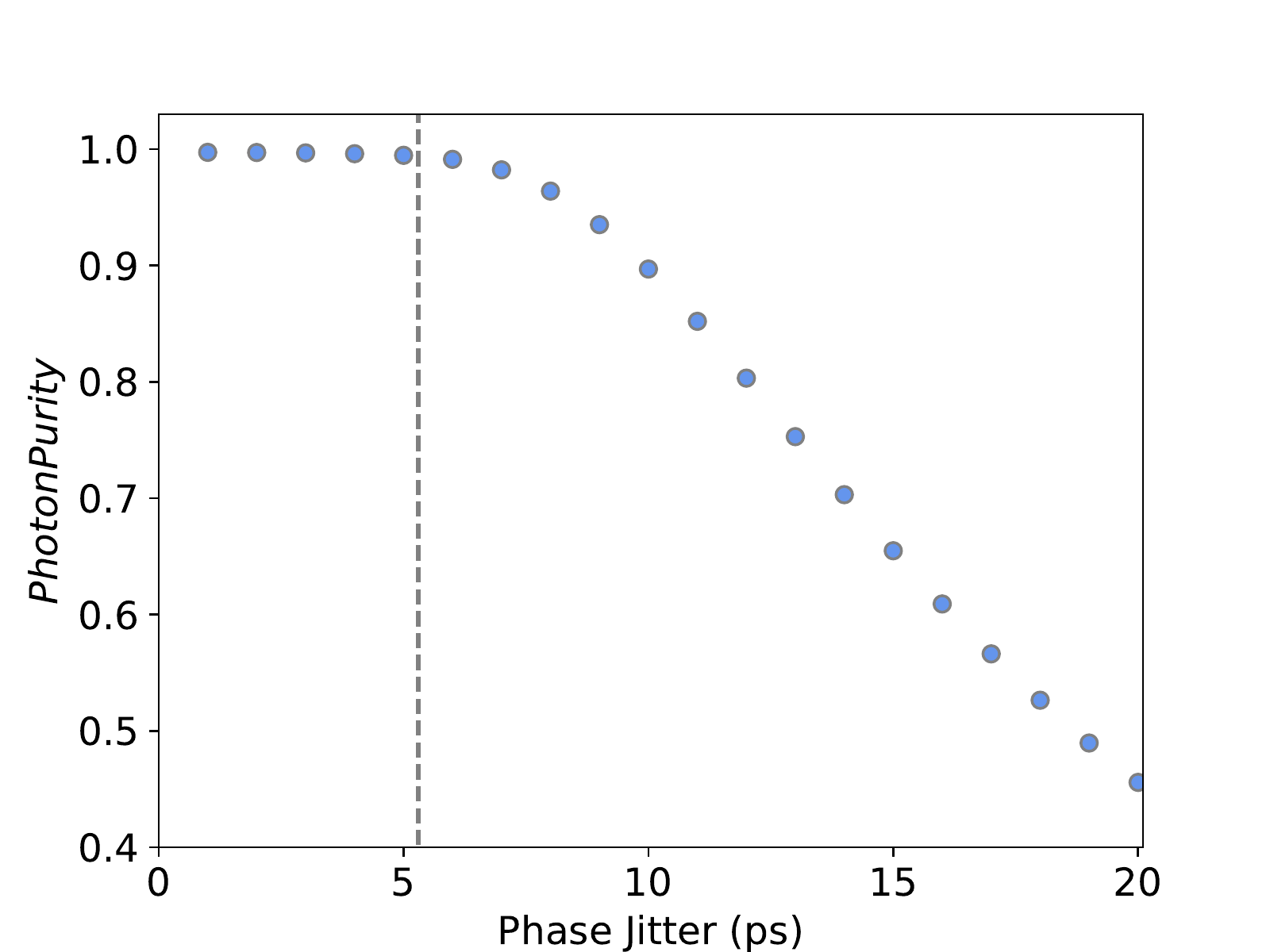}
		\caption{Simulated photon purity as a function of phase jitter. We assume perfect frequency resolution on the herald measurement and zero GVD. }
        \label{jitter}

\end{figure}

The simulated heralded photon purity as a function of the phase jitter $\sigma_{jitter}$ is shown in Fig \ref{jitter}. We estimate the phase jitter in our experiment to be $\sigma_{jitter} = 5.3$\,ps. Although we simulate that this jitter should not significantly reduce the photon purity, we note that a small and plausible increase in the phase jitter would cause a significant effect.

\section{Hong-Ou-Mandel Interference Visibility}

We measure the HOM interference of photons from subsequent time bins generated by the frequency multiplexed source. The measured visibility is $0.61 \pm 0.04$. This visibility is partly reduced due to the single-photon purity, discussed above, which it cannot exceed.

%
%
%
%

The HOM visibility is also reduced due to higher photon number contributions. These are estimated from the heralded second order correlation function at the pump power at which the two photon interference was measured, $g^{(2)}_\textrm{H} = 0.14$. If the source were single-mode, and the photons generated in pure states, the resulting HOM interference visibility would be $0.86$. With both the modeled purity of $0.84$ and multiphoton contributions considered, the expected HOM visibility is $0.72$. 

\section{ Losses}

The heralding efficiency, and therefore P(S,H) are significantly influenced by optical losses. The losses in our experiment are summarised in table \ref{component losses}. These values imply signal and herald efficiencies of $\eta_s = 0.13 $ and $\eta_H = 0.12$ (assuming the worst measured detector effciencies). We also infer the signal and herald efficiencies via measurement of the Klyshko efficiency. From this we obtain $\eta_s = 0.14 $ and $\eta_H = 0.11 $ to $0.15$. 

\begin{table}[h]
\begin{tabular}{c|c}  

\hline
Component   & Loss (dB)   \\

\hline \hline
SNSPDs    &      0.81-1.08     \\
Fibre Coupling Loss: Herald       &  3    \\
Fibre Coupling Loss: Signal       & 1.5       \\
Delay Line Insertion Loss       &     0.18       \\
Phase Modulator Insertion loss  &     2.2    \\
FBG Insertion Loss (TOF)  &  4.6  \\
Signal Filter Insertion Loss  &  0.46  \\
KTP Chip Loss  &  0.82  \\
Signal Filter Loss (Photon to Filter Bandwidth) & 3 \\
 & \\
\hline
Total Efficiency (inferred from components) &   Efficiency \\
\hline \hline
Signal   &   0.13 \\
Herald   &   0.12  \\
 
\hline
Klyshko Measurements  & Efficiency   \\
\hline \hline
$\eta_s$  & 0.14  \\
$\eta_h$  & 0.11 - 0.15    \\

\end{tabular}

\caption{Summary of component losses and efficiencies inferred from measurement of the Klyshko efficiency. }
\label{component losses}
\end{table}

Ultimately, the probability for the multiplexed source delivering a single photon will saturate at the signal arm efficiency (in the limit of infinite effective photon sources). The herald efficiency does not change maximum possible delivery probability. However a decrease in the herald arm efficiency increases the number of effective sources required in order to reach some fraction of the maximum single photon delivery probability. 

\begin{figure}[htb!]
	\centering
		\includegraphics[width=0.5 \textwidth]{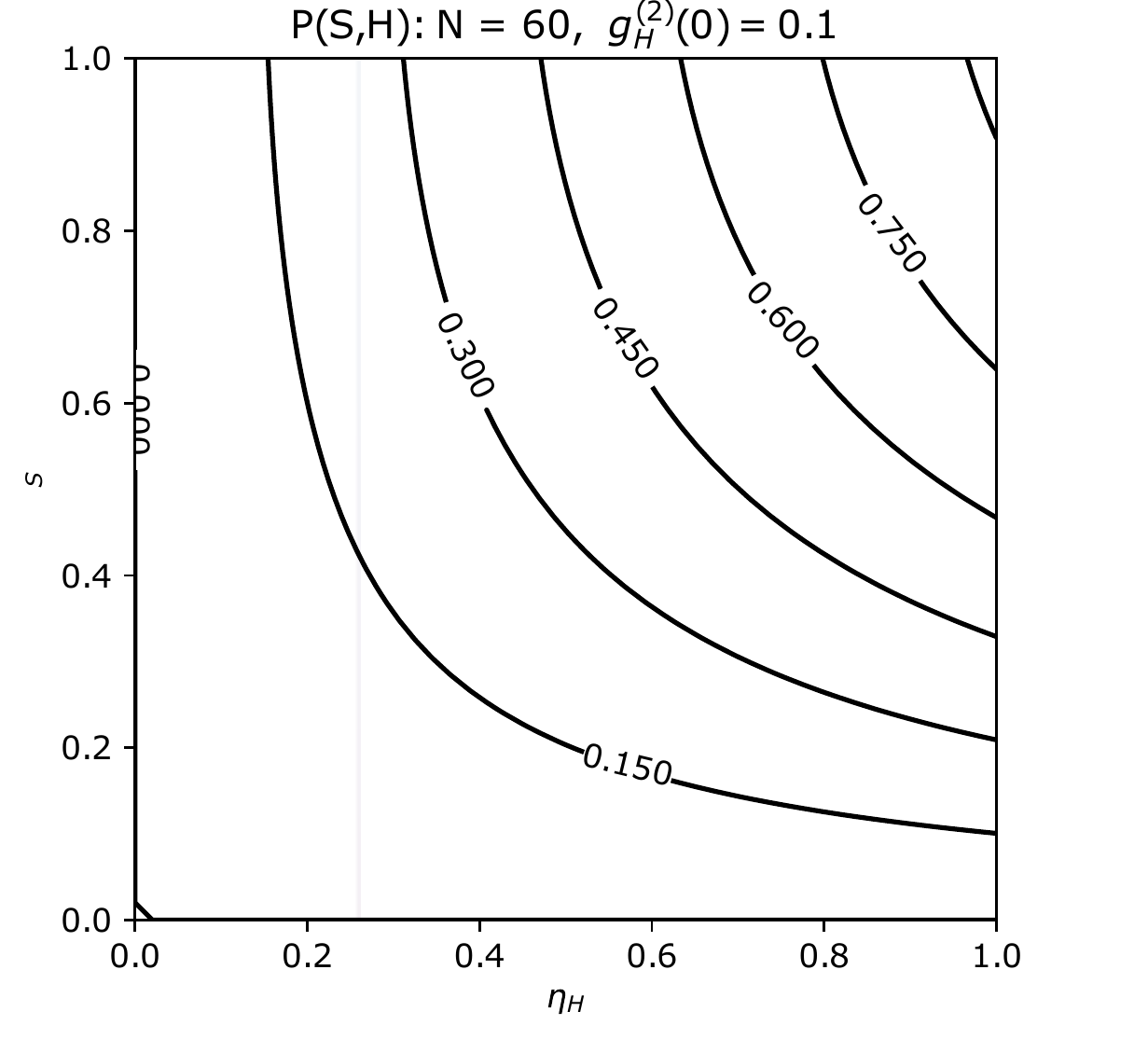}
		\caption{The joint probability of detection P(S,H) as a function of $\eta_S$ and $\eta_H$ for a fixed source number N and $g^{(2)}_H(0)$. }
        \label{stats}

\end{figure} 
To investigate the dependence of photon statistics on efficiencies, we model the joint spectrum as a series of factorable frequency modes, each of which has thermal statistics. The number of frequency modes, or effective sources being multiplexed is approximately $\Delta \omega$/$B$ where $\Delta \omega$ is the frequency shift range and $B$ is the single photon bandwidth.
In Figure \ref{stats} we plot the joint probability of detection P(S,H) for this model for a fixed effective source number and $g^{(2)}_H(0)$.

\bibliography{supp.bib}
